# Effective bowel motion reduction in mouse abdominal MRI using hyoscine butylbromide

Carlos Bilreiro[1,2,3]*, Francisca F. Fernandes[1]*, Luísa Andrade[2], Cristina Chavarrías[1], Rui V. Simões[1], Celso Matos[1,2], Noam Shemesh[1†].

[1]Champalimaud Research, Champalimaud Centre for the Unknown, Lisbon, Portugal.

[2]Radiology Department, Champalimaud Clinical Centre, Lisbon, Portugal.

[3]Nova Medical School, Lisbon, Portugal.

[†]Corresponding author:

Dr. Noam Shemesh, Champalimaud Research, Champalimaud Centre for the Unknown

Av. Brasilia 1400-038, Lisbon, Portugal.

E-mail: noam.shemesh@neuro.fchampalimaud.org ;

Phone number: +351 210 480 000 ext. #4467.

Word count: 2779

Running Head: High-resolution mouse abdominal MRI using hyoscine butylbromide administration.

*Carlos Bilreiro and Francisca F. Fernandes contributed equally to this work.

# ABSTRACT

**Purpose**: Bowel motion is a significant source of artifacts in mouse abdominal MRI. Fasting and administration of hyoscine butylbromide (BUSC) have been proposed for bowel motion reduction, but with inconsistent results and limited efficacy assessments. Here, we evaluate these regimes for mouse abdominal MRI at high field.

**Methods:** Thirty-two adult C57BL/6J mice were imaged on a 9.4T scanner with a FLASH sequence, acquired over 90 minutes with ~19s temporal resolution. During MRI acquisition, eight mice were injected with a low-dose and eight mice with a high-dose bolus of BUSC (0.5 and 5 mg/kg, respectively). Eight mice were food deprived for 4.5-6.5h before MRI and another group of 8 mice was injected with saline during MRI acquisition. Two expert readers reviewed the images and classified bowel motion, and quantitative voxel-wise analyses were performed for identification of moving regions. After defining the most effective protocol, high-resolution $T_2$-weighted and diffusion-weighted images were acquired from four mice.

**Results:** High-dose BUSC was the most effective protocol for bowel motion reduction, for up to 45 minutes. Fasting and saline protocols were not effective in suppressing bowel motion. High-resolution abdominal MRI clearly demonstrated improved image quality and ADC quantification with the high-dose BUSC protocol.

**Conclusion:** Our data show that BUSC administration is advantageous for abdominal MRI in the mouse. Specifically, it endows significant bowel motion reduction, with relatively short onset timings after injection (~8.5 minutes) and relatively long duration of the effect (~45 minutes). These features improve the quality of high-resolution images of the mouse abdomen.

**INTRODUCTION**

Abdominal MRI in the mouse is challenged by physiological motion, including breathing and bowel motility, causing multiple artifacts such as blurring and ghosting (1). Respiratory motion compensation is routinely achieved by techniques including respiratory triggering or placing animals in supine position (2,3). However, despite the importance of suppressing bowel motion, there is no consensus on how to achieve it in the mouse (4–8). Fasting and administration of hyoscine butylbromide (BUSC), trademarked Buscopan®, have been proposed, but the efficacy of these methods has never been thoroughly assessed. Food deprivation of at least 4 hours was reported to decrease peristalsis due to the absence of digestion and gut relaxation (4,7). Moreover, BUSC is a spasmolytic agent routinely used in the clinical setting to reduce motion artifacts from intestinal peristaltic motion (9–13). In the mouse, however, a consensus does not exist on its efficacy, and reported dosage regimens vary by several orders of magnitude, specifically from 0.5 to ≈80-100 mg/kg (4–6,8). Furthermore, the motion reduction window with BUSC was reportedly very short (≈5-10 minutes) (4,7), and some authors suggest that only a partial suppression of bowel movements can be achieved (6). Clearly, suppressing bowel motion over longer periods can be imperative for high-quality abdominal MRI. The aim of this study was thus to define a protocol for consistent bowel motion reduction (BMR) in the mouse and comprehensively evaluate its efficacy.

**METHODS**

All experiments were in accordance with European Directive 2010/63 and preapproved by the Institution's Review Board and the national competent authority. Our workflow is summarized in Figure 1, which included 4 animal groups with different BMR protocols: low- and high-dose

BUSC (4,8), food deprivation and saline. BMR was assessed using dynamic MRI of the abdomen in 32 mice, followed by high-resolution MRI with the most effective protocol in 4 animals. Below, we elaborate on each phase. Animal weights and ages are reported as mean ±standard deviation.

**Animal preparation**

Wild-type male mice (n=36) on a C57BL/6J background weighing 27.8 ±2.3 g and aged 14.8 ±4.6 weeks were used. Details on animal rearing and isoflurane-induced anesthesia can be found in Supplementary Information.

A 24G ×¾" catheter was inserted intraperitoneally in n=24 animals for BUSC or saline administration for characterizing the different BMR protocols, and in n=4 extra animals for high-resolution imaging. All volumes were injected over 10s using a syringe pump (GenieTouch$^{TM}$, Kent Scientific, Connecticut, USA), resulting in a 12-18 µL/s injection rate.

**MRI setup**

Animals were imaged using a 9.4T BioSpec® MRI scanner (Bruker, Karlsruhe, Germany) equipped with an AVANCE$^{TM}$ III HD console, producing isotropic pulsed field gradients of up to 660 mT/m (120 µs rise time), and a 40 mm-ID linear transmit-receive volume coil. All acquisitions began with routine adjustments: center frequency, RF calibration, acquisition of $B_0$ maps and automatic shimming.

**Characterization of BMR**

MRI protocol

A FLASH sequence with two 1-mm thick coronal slices positioned in the abdomen and ~19s temporal resolution (detailed in Supplementary Information) was acquired in n=32 animals.

BMR protocols

Mice were divided into four groups. In the Food Deprivation group, 8 animals (16.5 ±2.1 weeks-old, 28.7 ±1.3 g) were fasted between 4h30min to 6h30min prior to the experiment. In the Low-Dose BUSC group, 8 animals (15.3 ±3.1 weeks-old, 28.1 ±1.9 g) were injected with a 5 mL/kg bolus of 0.5 mg/kg (4) BUSC (Buscopan®, Boehringer Ingelheim, Barcelona, Spain: 20 mg/ml, diluted 1:200 in saline) after 10 min of acquisition. In the High-Dose BUSC group, 8 animals (12.4 ±4.2 weeks-old, 27.9 ±3.8 g) were injected with a 5 mL/kg bolus of 5 mg/kg (8) BUSC (1:20 dilution) after 10 min of acquisition. Finally, in the Saline control group, 8 animals (12.7 ±6.0 weeks-old, 26.5 ±0.8 g) received an injection of 5 mL/kg saline after 10 min of acquisition.

Data analysis

The data analysis included two main parts: a qualitative analysis of the images, performed by two readers, aiming to identify and define periods with BMR; and a quantitative analysis of the data, aiming to produce motion maps and time-courses.

   A. Qualitative analysis

Two gastrointestinal radiologists (7 and 9 years of experience) reviewed the acquired images, blinded to the interventions performed on the animals, and separately divided the time-course of every slice into time-blocks with different degrees of peristaltic motion: 1 - strongly reduced peristalsis; 2 - partial reduction of peristalsis with residual movement of bowel content; 3 - persistent peristalsis. Blocks with a classification of 1 and 2 were then merged to define periods with BMR and obtain two relevant timings: the time from the beginning of acquisition to the start of BMR ($T_{mr}$); and the duration of BMR ($D_{mr}$).

To evaluate inter-rater variability, $T_{mr}$ and $D_{mr}$ were compared between readers with single score intra-class correlation coefficient (ICC) assessing absolute agreement and using a two-way random effects model, using SPSS® (SPSS Inc., NY, USA). In cases where bowel motion was maintained throughout the entire scan (classification = always 3), values of 90 min and 0 min were assigned to $T_{mr}$ and $D_{mr}$, respectively.

Finally, both readers defined in consensus time blocks and classifications for each animal, and $T_{mr}$ and $D_{mr}$ were recalculated. The duration of halted motion ($D_{mh}$), i.e. the time under a classification of 1, was also calculated. For the low- and high-dose BUSC groups, the time from the injection of BUSC to the start of BMR ($T_{mrb}$) and the duration of BUSC effects ($D_{mrb}$) were also computed. $T_{mrb}$ was calculated by subtracting 10 min from $T_{mr}$, since BUSC was injected after 10 min of acquisition, and $D_{mrb} = D_{mr}$. Mann-Whitney test was used to evaluate differences between BUSC protocols on $T_{mrb}$ and $D_{mrb}$ and Kruskal-Wallis test for differences between the four groups on $D_{mh}$ (significance at 0.05).

B. Quantitative Analysis

Datasets were pre-processed (details in Supplementary Information) and analyzed in MATLAB® (MathWorks Inc., Natick, MA). The absolute differences between consecutive images were computed voxel-by-voxel, summed for 10-11 min intervals throughout the acquisition and mapped voxel-wise for identification of moving regions. These absolute differences were thresholded at 20% of the highest difference obtained from all scans for identification of significantly moving pixels at each time instant, and the differences from the pixels surviving the thresholding were summed to create motion time-courses for each animal. Individual motion time-courses were then averaged for each group to generate group time-courses.

To investigate anesthesia depth and temperature as possible confounding factors, influencing BMR and contributing to the differences observed between groups in the 20-50 min period after the beginning of MRI, respiratory rate and rectal temperature values averaged for each animal in that period were tested with ANOVA (significance at 0.05).

**High-resolution MRI**

After determining the most effective protocol for BMR, we chose to demonstrate its value with high-resolution MRI. $T_2$-weighted and diffusion-weighted images were acquired in the abdomen of n=4 mice, before and ~6.5 min after the administration of high-dose BUSC, using RARE and EPI-based Stejskal-Tanner DWI, respectively (detailed in Supplementary Information). ADC maps were calculated using MATLAB®. ImageJ (US NIH) was used for manual drawing of regions of interest (ROIs) for comparing results between ADC measurements.

**RESULTS**

<u>Agreement between readers</u>. Video S1 shows the FLASH images obtained throughout 90-min scans from one representative animal of each experimental group, revealing periods with different degrees of motion. Figure 2 shows different degrees of peristaltic motion, assigned by the readers to time blocks of each animal. Reliability analysis (Figure S2) revealed a very high agreement between readers for $T_{mr}$ (ICC(2,1)=0.984) and $D_{mr}$ (ICC(2,1)=0.951).

<u>Readers' consensus on the efficacy of each protocol</u>. Table S1 shows $T_{mr}$, $D_{mr}$ and $D_{mh}$ for each animal. Most mice in the food deprivation and saline groups never showed significant BMR. In contrast, animals under low- or high-dose BUSC showed sustained BMR after injection. In particular, $T_{mrb}$ and $D_{mrb}$ in the low-dose BUSC group were 12.9 (±9.6) min and 46.8 (±18.3) min, respectively. $T_{mrb}$ was less variable in the high-dose BUSC group (8.5 ±2.8 min), whereas $D_{mrb}$ was similar (40.8 ±17.8 min), without significant differences ($T_{mrb}$, $P=0.59$; and $D_{mrb}$, $P=0.33$). Nevertheless, half of the mice under high-dose BUSC showed absent peristalsis, while this was observed in only one animal in the low-dose BUSC and in none of the animals from other groups. Most animals (7 out of 8) under low-dose BUSC showed a reduced but residual peristalsis. Kruskal-Wallis test revealed significant differences on $D_{mh}$ ($P<0.03$) between the four groups. After pairwise comparisons, significant differences were only found between the high-dose BUSC and both food deprivation and saline groups (both $P=0.04$). No significant differences were observed between the low-dose and the high-dose BUSC, saline or food deprivation groups ($P=0.25$, $P=0.87$ and $P=0.87$, respectively).

<u>Motion maps and time-courses</u>. The quantitative analysis of bowel motion was mostly consistent with the readers' analysis. Figure 3 shows strongly reduced bowel motion in one representative animal from the high-dose BUSC group (from 18-28 min to 53-63 min), an effect noticeably more intense compared to the others. This effect was reproducible throughout the entire high-dose BUSC group (Figure S3) and in the individual and group motion time-courses (Figures S4 and 4, respectively). The animal-averaged time-course from the high-dose BUSC group shows a robust decrease in motion for the longest period of time in comparison to the others. A maintained peristalsis is observed throughout the entire acquisition in all food deprivation and most saline groups' animals. Notably, a sudden increase in motion was consistent in both BUSC and saline groups during the first five minutes after injection (Figures 4, S4).

Animals' welfare. No side effects of BUSC administration were observed for either dosages, and all animals quickly recovered from sedation. Group-averaged respiratory rate and rectal temperature values recorded during MRI acquisition are plotted in Figure S5. ANOVA did not reveal significant differences between groups (respiratory rate, $P$=0.28; rectal temperature, $P$=0.39).

High-resolution MRI. The introduction of high-dose BUSC clearly improved image quality regarding structural definition, organ detail and reduction of motion artifacts. These effects are noticeable in both $T_2$-weighted RARE (Figures 5A, S6-7) and DWI (Figures 5B, S8-9), when comparing depiction of bowel walls and parenchymal organs (kidneys, pancreas, spleen, liver). Regarding DWI, BUSC effect is mostly noted in b1000 images, representing a powder average of 14 directions, acquired over ~30 minutes. BUSC provided reduced data variability derived from motion artifacts in ADC measurements from ROIs in the kidney parenchyma and renal pelvis (Figures 5B, S10).

**DISCUSSION**

Our study reveals that BUSC is effective for BMR in the mouse for an average duration of 40-45 min. Its effects were consistent in both qualitative and quantitative analyses, with reduced bowel motion using both BUSC dosages. However, strongly reduced peristalsis was best achieved with the high-dose protocol, significantly improving abdominal imaging quality by reducing motion artifacts. This study provides the first thorough characterization of BMR strategies for abdominal MRI in the mouse.

These results reveal a longer duration of BUSC effects on mouse bowel peristalsis than previously reported by Grimm et al. (4). Additionally, fasting, suggested by others (4,7) to have comparable effects to BUSC, was not effective (Figure 3). These differences likely result from our motion quantification strategy involving both subjective and objective criteria, whereas previous studies did not report such quantification of bowel motion. BUSC works as a blocker of muscarinic receptors for acetylcholine in smooth muscle cells in the gastrointestinal tract, reducing motility in a dose-dependent manner (14). This effect is evidently more intense and reliable than the motility reduction obtained with fasting.

Our motion quantification strategy did not always correlate with the readers' evaluation. Discrepancies between both methods are apparent in the time-course graphs (Figure S4). The readers' evaluation was more specific in detecting peristaltic motion, as the quantitative analysis mostly reflects the magnitude of motion in the entire FOV. Therefore, it does not strikingly reflect the presence of motion when it happens in a small number of voxels or if it does not translate into large differences in voxel intensity (e.g. Figure S4: Animal #29). Also, susceptibility artifacts caused by bowel gas are very noticeable in the quantitative analysis, creating "spikes" in the time-course that do not actually correspond to bowel motion. These can be recognized by human readers as artifacts and classified as residual motion of bowel content (e.g. Figure S4: Animals #7, #16, #27, #28). Thus, both strategies are complementary: the qualitative focuses on the detection and classification of peristaltic motion and the quantitative represents objectively measured data, highly sensitive to motion, allowing its representation over time. Moreover, it should be noted that these case-by-case discrepancies were partially attenuated by averaging (Figure 4).

Considering the motion time-courses, a shorter magnitude of motion can be observed in the food deprivation group, before injections occurred. This, however, did not translate into significant BMR in the readers' evaluation. Nevertheless, it might be hypothesized that a combination of food deprivation and BUSC administration, especially at higher dosages, could provide even greater BMR. Since the high-dose BUSC protocol provided satisfactory results without side effects, this combination might not be worth the additional time and preparation.

Isoflurane-induced anesthesia has been reported to suppress gastrointestinal motility in rodents (15,16). In this study, the anesthesia depth did not vary substantially between groups, as the respiratory rate and rectal temperature values recorded during MRI acquisition (Figure S5) were on the same ranges, including the period with stronger BUSC effects (20-50 minutes after beginning of acquisition). Therefore, the use of isoflurane should not have contributed to the differences observed in bowel motion.

The high-resolution data acquired under high-dose BUSC reveals well-defined and detailed abdominal organs such as the bowel, kidneys, pancreas, spleen and liver (Figures 5, S6-9). Previous studies reported $T_2$-weighted TSE abdominal images with in-plane resolutions as high as 75x156 $\mu m^2$ (1 mm slice thickness) (6), 117x70 $\mu m^2$ (0.5 mm slice thickness) (4) and 100x100 $\mu m^2$ (1 mm slice thickness) (17). Our results show that higher resolutions (85x85 $\mu m^2$) with reduced slice thickness (0.3 mm) can be achieved without motion artifacts when using high-dose BUSC. Similarly, while previously reported DWI resolutions reached 1300x1300 $\mu m^2$ (2 mm slice thickness) (18), 250x250 $\mu m^2$ (1 mm slice thickness) (19), 700x700 $\mu m^2$ (1.5 mm slice thickness) (20) and 1250x1250 $\mu m^2$ (1.3 mm slice thickness) (21), we achieved a significantly higher resolution (150x150 $\mu m^2$; 0.65 mm slice thickness) free of motion artifacts.

These results favor the use of BUSC for long acquisitions using several repetitions/averages for producing a single image, exemplified by the DW images presented. The longer the acquisition time, the longer the data will be exposed to the occurrence of motion and, as each repetition is uniquely affected by motion, the resulting artifacts increase when a powder-average of several diffusion directions is used to create DW images. Moreover, ADC maps or other quantifiable measurements are expected to be more accurately calculated with the reduction of motion-dependent artifacts (22,23). Our results reflect this, with more precise ADC measurements in the kidney using high-dose BUSC (Figures 5, S10), demonstrating how BUSC administration is useful for motion-sensitive organs (kidneys, bowels, pancreas) in quantitative MRI.

As established for humans, glucagon might be an alternative to BUSC, a hormone known to have a similar, although shorter, effect (24). Further studies should investigate its use as an alternative to BUSC in the mouse.

As in every work, we acknowledge several relevant limitations in this study. First, the statistical evaluation was limited, probably due to insufficient power: trends were observed favoring the high-dose compared with the low-dose BUSC protocol for more reliable initiation and duration of BMR, without reaching statistical significance. Nevertheless, significant differences were observed between the high-dose BUSC and the food deprivation/saline groups on $D_{mh}$, while the low-dose BUSC did not show significant differences with any other groups. Second, the injections performed at room temperature (~20 ºC) might have caused the short-duration motion increases observed immediately after all injections. This should be taken into account when considering repeated bolus protocols for long periods of BMR, as the motion caused by each injection could be detrimental to data acquisitions. In this context, a-priori warming the solution to body temperature or using continuous infusion protocols might prove beneficial. Third, the only route for BUSC administration was intraperitoneal due to the ease of catheter placement; other routes of administration (intravenous or intramuscular) might provide different results (25).

Lastly, since no side effects were observed with our dosage protocols, future studies could investigate mice's capability to tolerate higher dosages and their efficacy on BMR.

## CONCLUSIONS

BUSC administered in a single 5 mg/kg bolus is effective for BMR in the mouse for up to 45 min, enabling robust imaging in the abdomen and allowing high-resolution imaging in this challenging region. Our results are promising for future studies in motion-susceptible organs (bowels, pancreas and kidneys), either for diagnostic or quantification purposes.

## ACKNOWLEDGEMENTS

Funding Support: Champalimaud Foundation; H2020-MSCA-IF-2018, ref: 844776.

CONGENTO LISBOA-01-0145-FEDER-022170.

22. le Bihan D, Poupon C, Amadon A, Lethimonnier F. Artifacts and pitfalls in diffusion MRI. Journal of Magnetic Resonance Imaging 2006;24:478–488 doi: 10.1002/jmri.20683.

23. Ozaki M, Inoue Y, Miyati T, et al. Motion artifact reduction of diffusion-weighted MRI of the liver: Use of velocity-compensated diffusion gradients combined with tetrahedral gradients. Journal of Magnetic Resonance Imaging 2013;37:172–178 doi: 10.1002/jmri.23796.

24. Froehlich JM, Daenzer M, von Weymarn C, Erturk SM, Zollikofer CL, Patak MA. Aperistaltic effect of hyoscine N-butylbromide versus glucagon on the small bowel assessed by magnetic resonance imaging. European Radiology 2009;19:1387–1393 doi: 10.1007/s00330-008-1293-2.

25. Gutzeit A, Binkert CA, Koh DM, et al. Evaluation of the anti-peristaltic effect of glucagon and hyoscine on the small bowel: Comparison of intravenous and intramuscular drug administration. European Radiology 2012;22:1186–1194 doi: 10.1007/s00330-011-2366-1.

**FIGURES/TABLES**

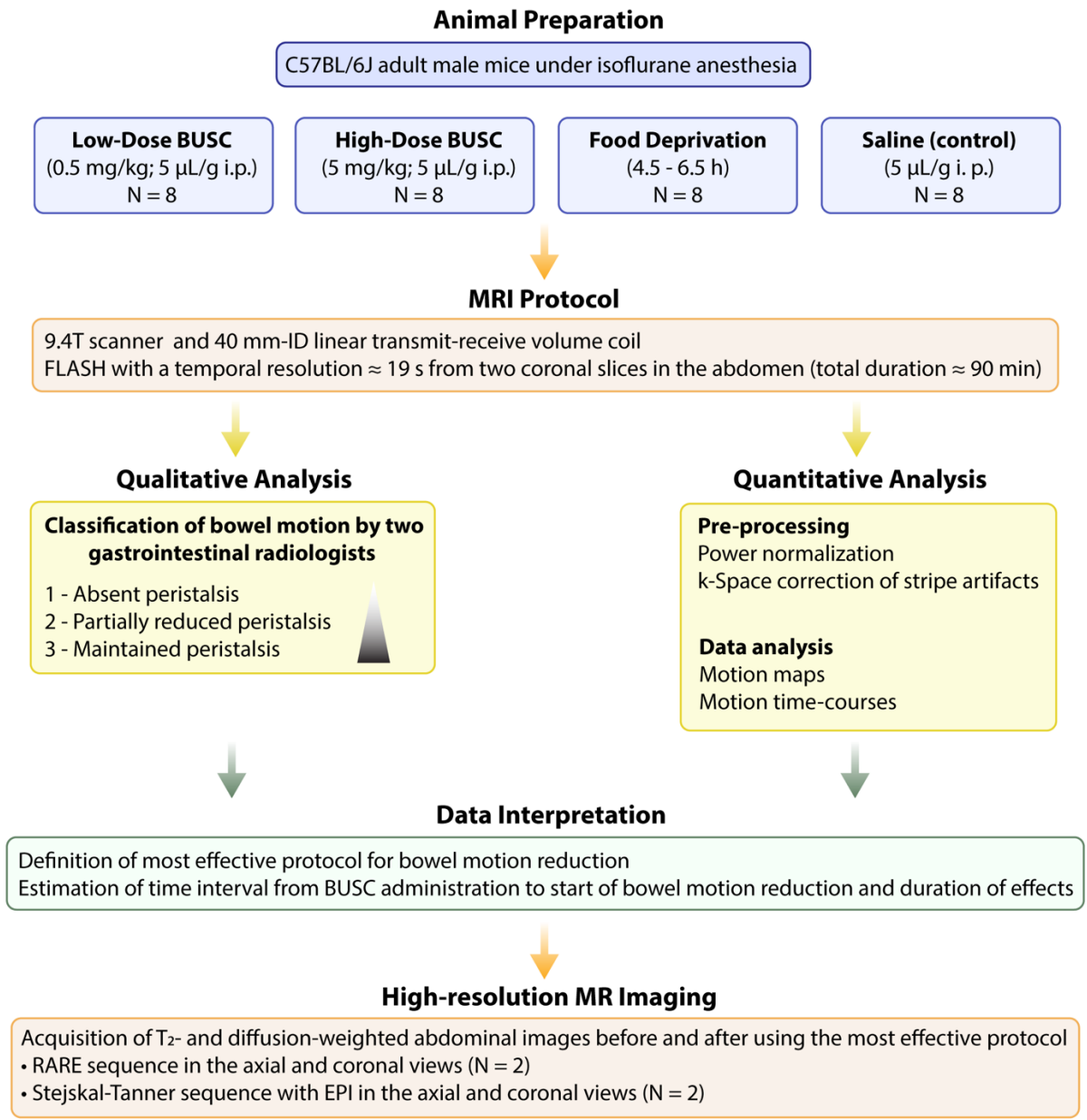

Figure 1 – **Study flowchart.** Thirty-two C57BL/6J mice were subjected to four different bowel motion reduction protocols and underwent MRI scans under isoflurane anesthesia. Images were qualitatively and quantitatively assessed for the presence and degree of bowel motion in each condition. After determining the most effective protocol, high-resolution MR images of the entire abdomen were acquired in four animals, before and after using that protocol.

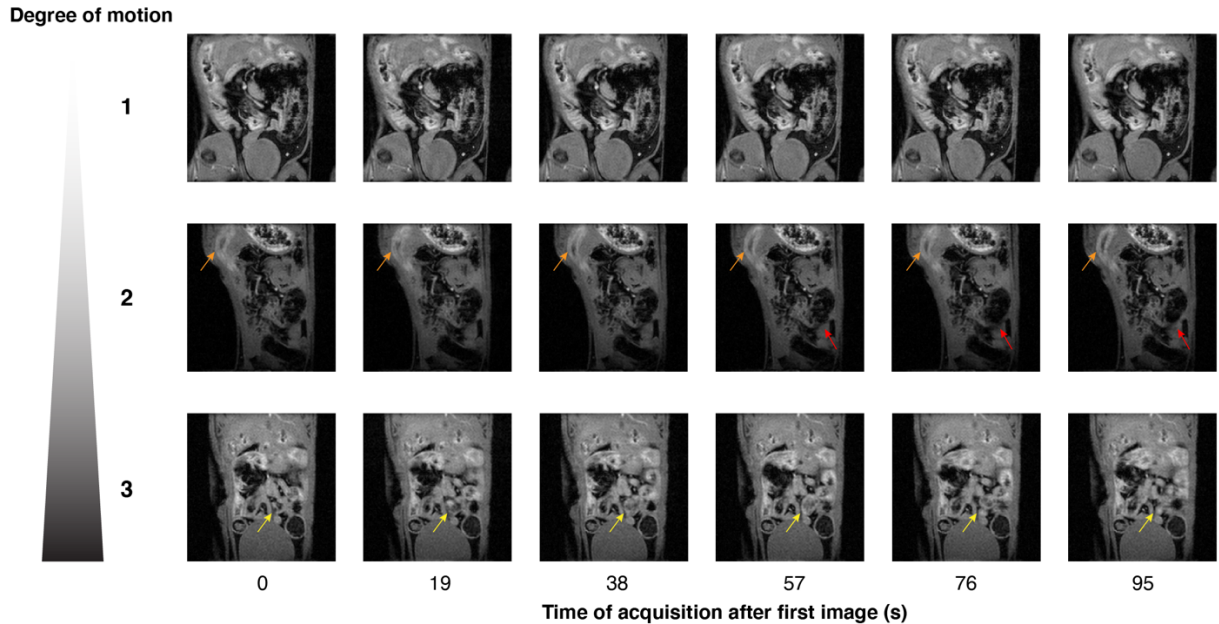

Figure 2 – **Classification of images according to the degree of motion.** Three examples of 6 consecutive images from the mouse abdomen acquired with a FLASH sequence in the coronal plane. Each row represents a time period of approximately 1.5 min. Upper row: No noticeable peristalsis is observed (classification = 1). Middle row: Residual motion of bowel content without peristaltic motion, causing intensity changes (orange arrows) and susceptibility artifacts (red arrows) (classification = 2). Lower row: Peristaltic bowel motion. Notice the variability of luminal diameter in one bowel loop along this time period (yellow arrows) (classification = 3).

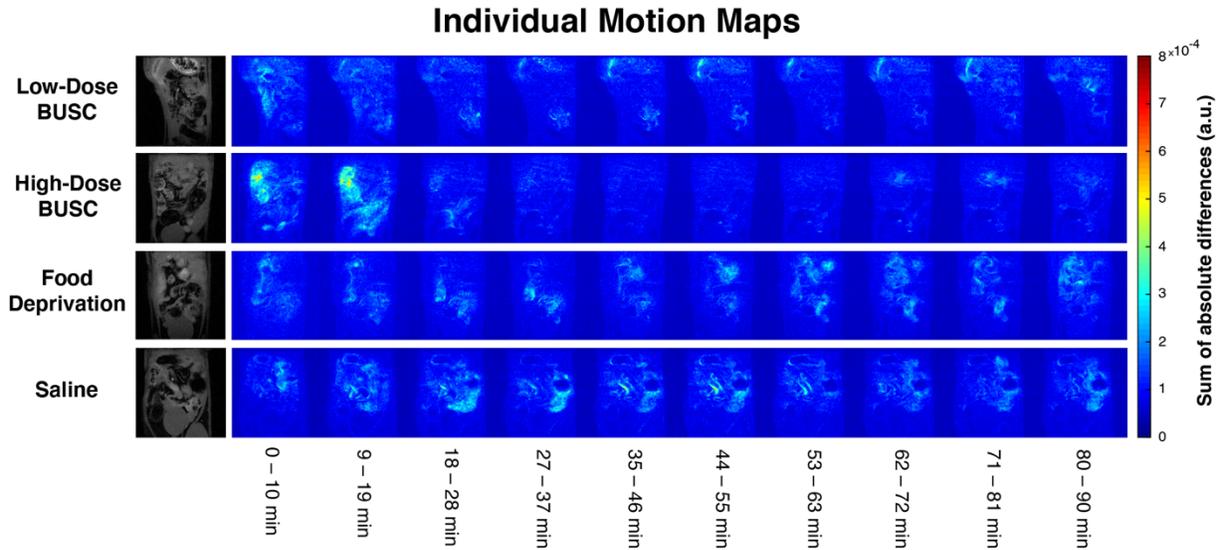

Figure 3 – **Representative individual motion maps**. Voxel-by-voxel motion maps obtained for one representative slice of one animal from each group (specifically, from Animals #3, #14, #17 and #30), representing the calculated sum of the absolute differences between consecutive images for 10-11 min intervals throughout the acquisition. Dark blue areas represent regions without significant bowel motility. A raw image from the slice is presented on the left of each map for anatomical region definition.

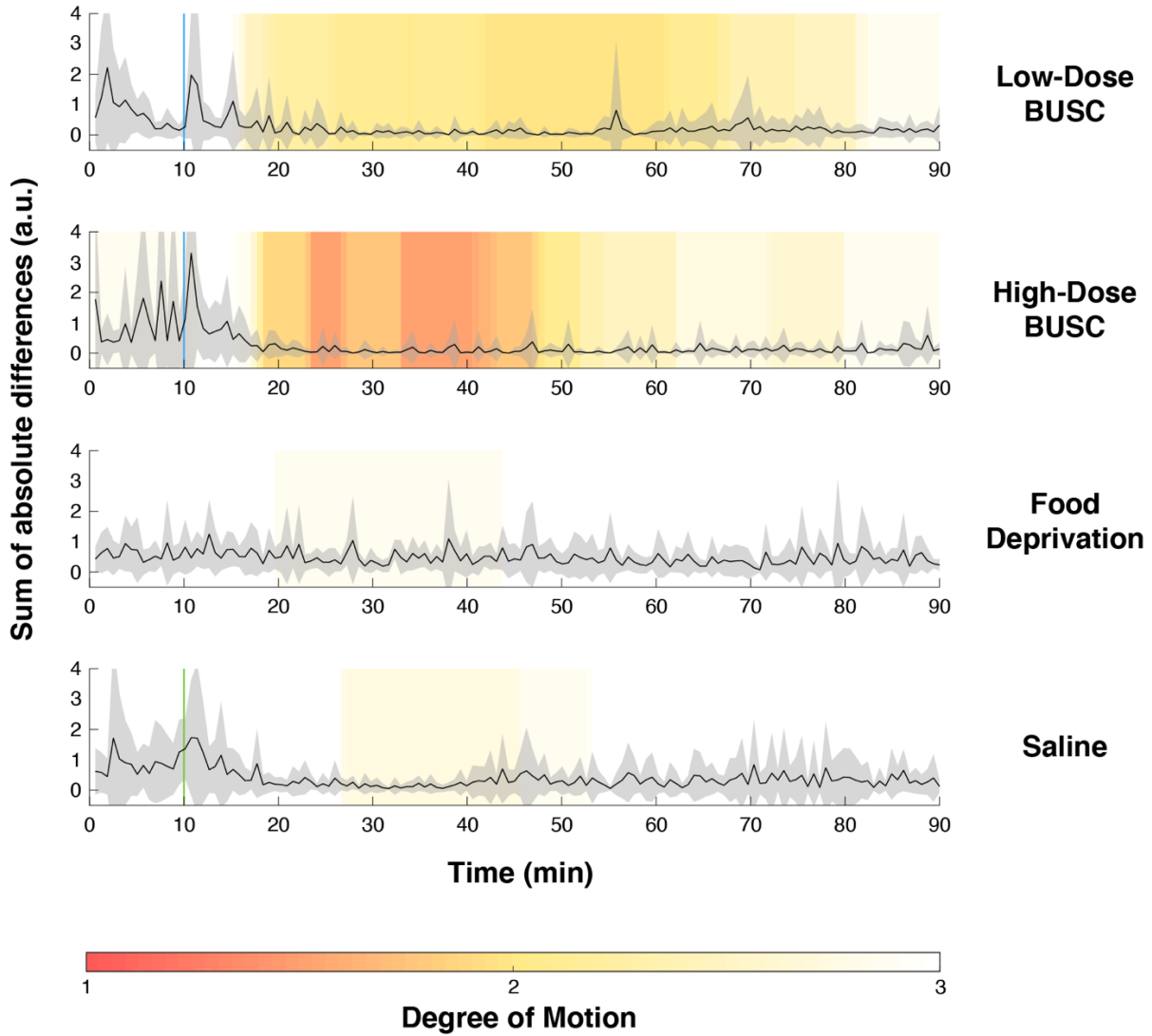

Figure 4 – **Group motion time-courses**. Estimated group-averaged motion time-courses obtained from both slices of all animals from each group, depicting the sum of absolute differences between consecutive images for ~38 sec intervals throughout the acquisition from significantly moving pixels, i.e. pixels whose absolute difference was higher than 20% of the highest difference, at each time instant. Gray shaded areas represent the standard deviation of the estimated motion. The degrees of motion defined in consensus by the two readers for each time interval of each animal

were averaged for each group and are color coded behind each time-course, where white represents maintained peristalsis (classification = 3), bright yellow represents reduced but residual peristalsis (classification = 2) and red represents strongly reduced peristalsis (classification = 1). Blue lines indicate i.p. injection of a BUSC bolus whereas the green line represents i.p. injection of saline.

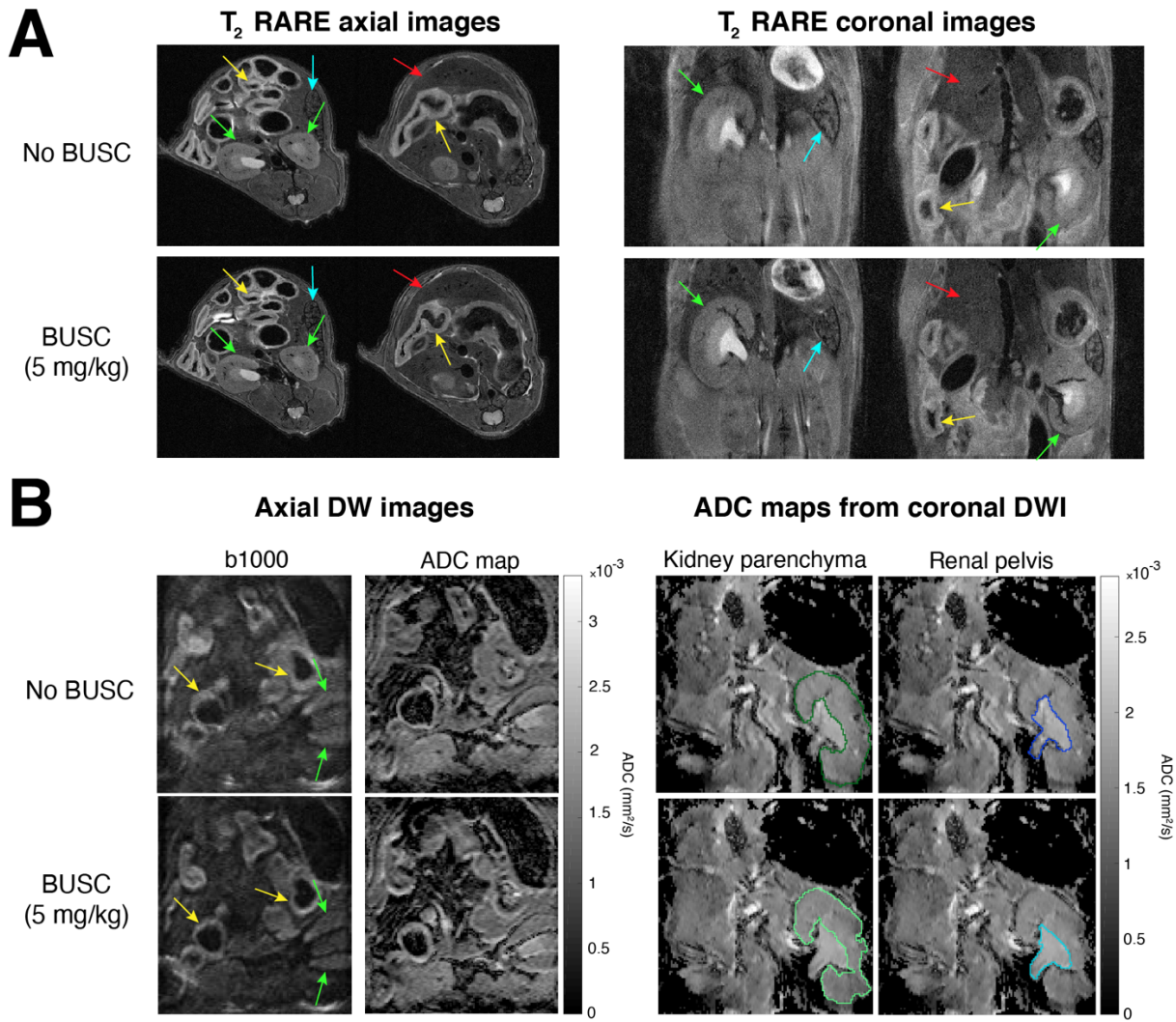

Figure 5 – **High-resolution T$_2$ RARE and diffusion-weighted images of the abdomen, with corresponding ADC maps, before and after high-dose BUSC injection**. Representative axial and coronal T$_2$-weighted (A) and diffusion-weighted (B) images of the mouse abdomen acquired without BUSC and 6.5 min after BUSC i.p. injection. Yellow arrows point to the bowel, green arrows to the kidneys, blue arrows to the spleen and red arrows to the liver. The ROIs performed for ADC values quantification in the renal pelvis and parenchyma are presented in (B), on the right, with blue and green contours respectively.

Video S1 – **Videos from each experimental group**. FLASH images obtained during 90 min from one representative slice of four animals (specifically, from Animals #3, #14, #17 and #30). The intraperitoneal injections at 10 minutes after the start of each scan are represented by a blue (BUSC) or gray (saline) circle in the left top corner of each video, when applicable. The low-dose BUSC scan shows a period with reduced but residual peristalsis after the injection, classified by the readers with a degree of motion of 2. The high-dose BUSC scan shows a strongly reduced peristalsis after the injection, whose period was classified as 1. At the end of both these scans, there is a return to normal peristaltic motion as the BUSC effects wear off. Both the food deprivation and saline scans show continuous motion throughout the entire scan and were therefore classified with a degree of motion of 3. All images displayed were corrected for transient global signal changes and stripe artifacts.

Table S1 – **Relevant timings defined in consensus for each animal**. Time from the beginning of acquisition to the start of noticeable bowel motion reduction ($T_{mr}$), as well as the duration of those effects ($D_{mr}$), i.e. the time under a classification of 1 or 2. The duration of halted motion ($D_{mh}$), i.e. the time under a classification of 1, is also included. For the low- and high-dose BUSC groups, $T_{mrb}$ was calculated by subtracting 10 min from $T_{mr}$, since BUSC bolus injection was performed after 10 min of acquisition, and $D_{mrb}$ was equal to $D_{mr}$ (not in the table). Since bowel motion reduction in Animal #9 started before the injection of BUSC, its $T_{mrb}$ and $D_{mrb}$ values were 7.1 min and 34.9 min, respectively. NR: No reduction.

| Group | Animal | Time to start motion reduction ($T_{mr}$, min) | Duration of motion reduction ($D_{mr}$, min) | Duration of halted motion ($D_{mh}$, min) |
|---|---|---|---|---|
| Low-Dose BUSC | 1 | 15.5 | 74.5 | 0.0 |
|  | 2 | 41.2 | 41.2 |  |
|  | 3 | 25.4 | 49.4 |  |
|  | 4 | 18.1 | 49.8 |  |
|  | 5 | 16.2 | 65.0 | 65.0 |
|  | 6 | 33.3 | 33.3 | 0.0 |
|  | 7 | 14.9 | 46.3 |  |
|  | 8 | 19.0 | 14.9 |  |
| High-Dose BUSC | 9 | 0.0 | 52.0 | 29.2 |
|  | 10 | 17.1 | 37.4 | 0.0 |
|  | 11 | 17.7 | 29.5 | 29.5 |
|  | 12 | 15.2 | 33.0 | 0.0 |
|  | 13 | 18.1 | 61.8 | 61.8 |
|  | 14 | 23.1 | 39.0 | 39.0 |
|  | 15 | 22.5 | 18.4 | 0.0 |
|  | 16 | 17.1 | 72.9 |  |
| Food Deprivation | 17 | NR | 0.0 | 0.0 |
|  | 18 |  |  |  |
|  | 19 |  |  |  |
|  | 20 |  |  |  |
|  | 21 |  |  |  |
|  | 22 | 19.3 | 24.4 |  |
|  | 23 | NR | 0.0 |  |
|  | 24 |  |  |  |
| Saline | 25 | NR | 0.0 |  |
|  | 26 |  |  |  |
|  | 27 | 26.3 | 19.3 |  |
|  | 28 | 26.6 | 26.3 |  |
|  | 29 | NR | 0.0 |  |
|  | 30 |  |  |  |
|  | 31 |  |  |  |
|  | 32 |  |  |  |

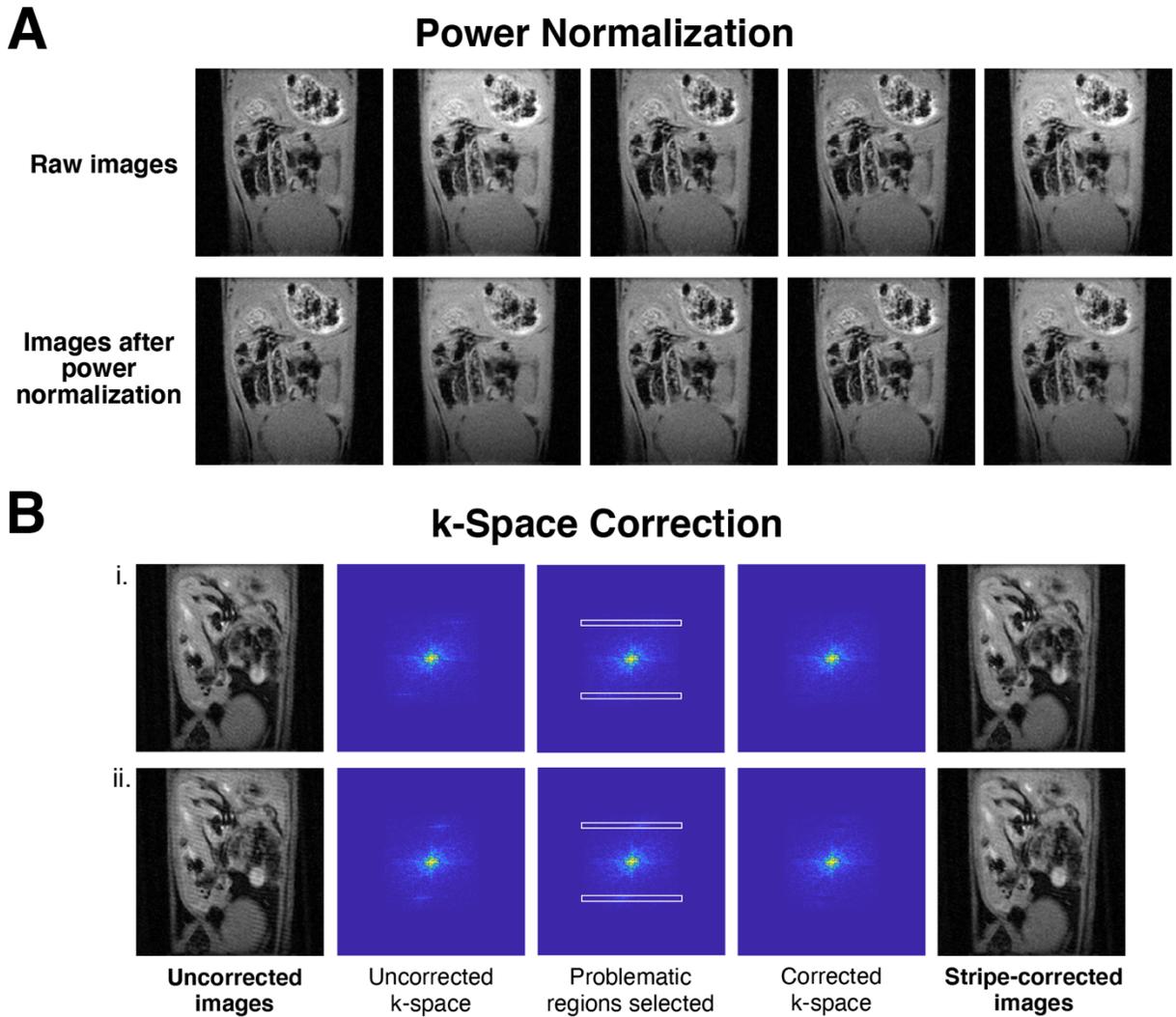

Figure S1 – **Pre-processing of abdominal MR images**. (A) Correction of transient global signal increases on 5 consecutive frames from a representative animal through power normalization, where power represents the sum of signal intensities from all voxels in each slice frame. Upper row: Raw images with different powers. Lower row: Images with unitary power after normalization, revealing a significant reduction of global signal variations. (B) Correction of stripe artifacts on 2 representative frames from a single animal through elimination of spurious signals

at high-frequency regions of the k-space. White boxes delineate regions prone to correction that were manually selected for that animal.

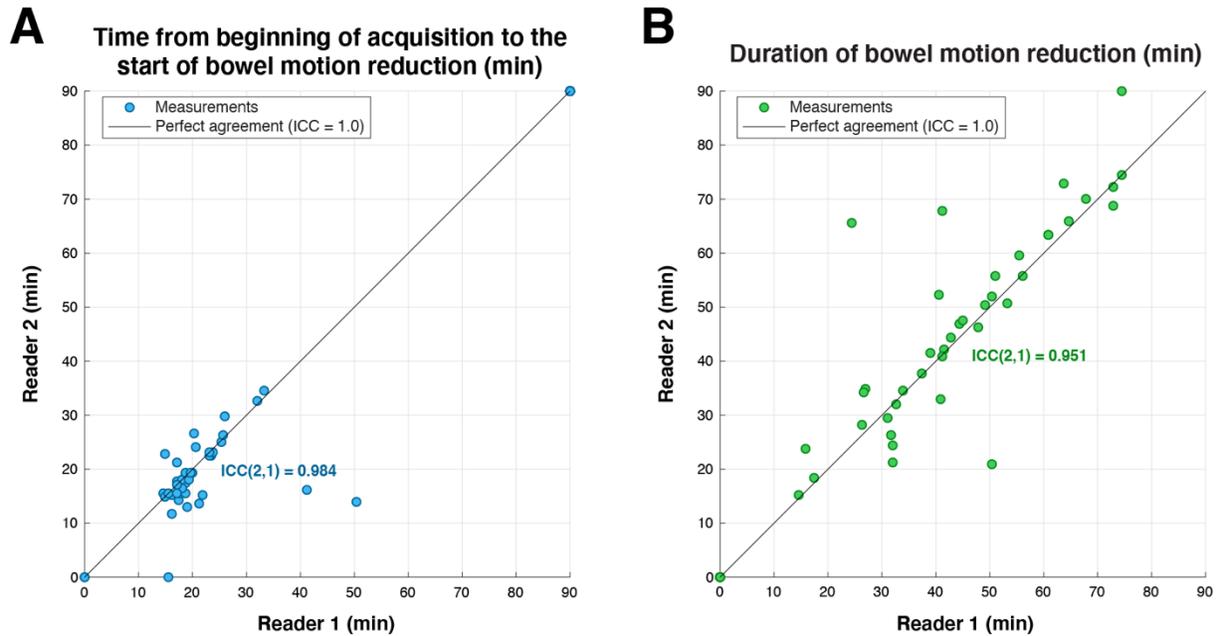

Figure S2 - **Reliability analysis**. (A) Time from the beginning of acquisition to the start of bowel motion reduction ($T_{mr}$) and (B) duration of small motion period ($D_{mr}$) measured by two readers for all animal slices (N = 64 slices) and respective measures of inter-rater agreement (ICC(2,1) = 0.984 and ICC(2,1) = 0.951). The calculated ICCs reveal a very high agreement between readers.

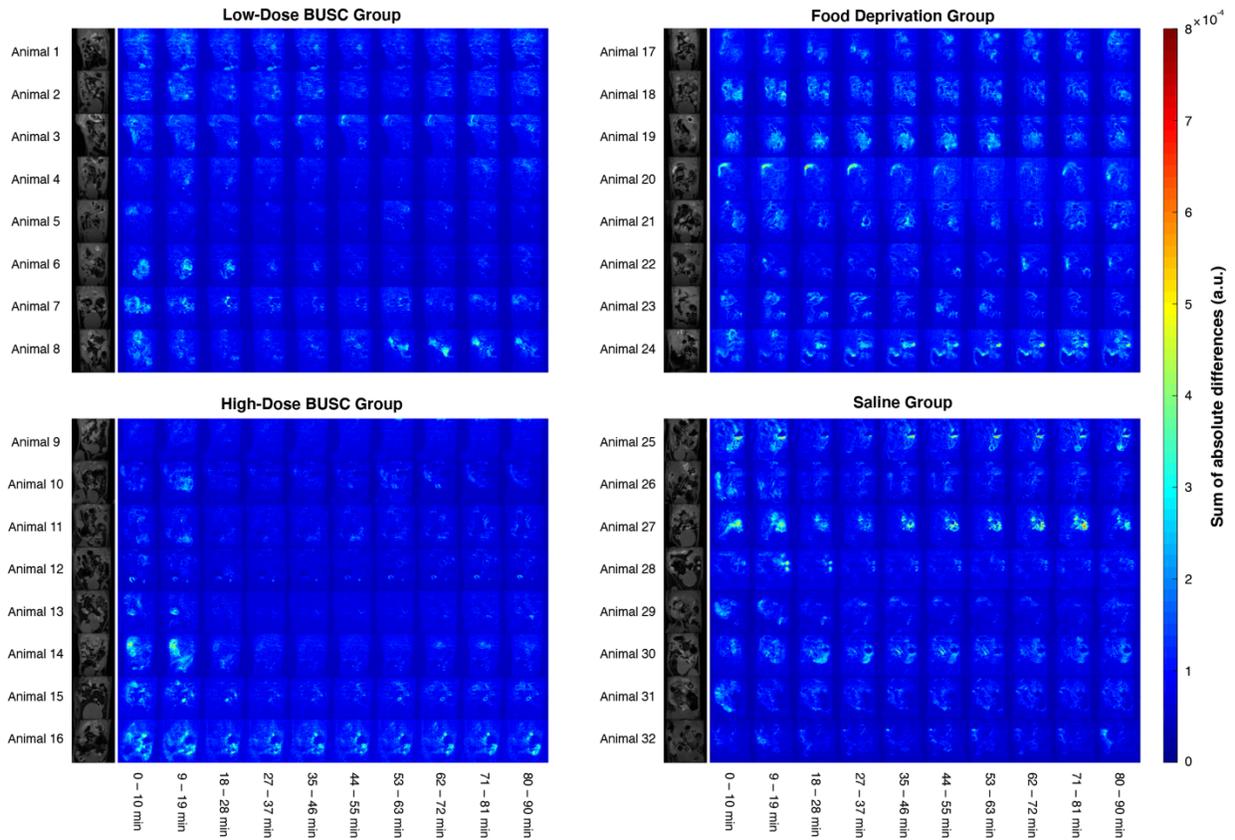

Figure S3 - **Individual motion maps**. Voxel-by-voxel motion maps obtained for one representative slice of each animal, representing the calculated sum of the absolute differences between consecutive images for 10-11 min intervals throughout the acquisition. Dark blue areas represent regions without significant bowel motility. A raw image from the slice is presented on the left of each map for anatomical region definition.

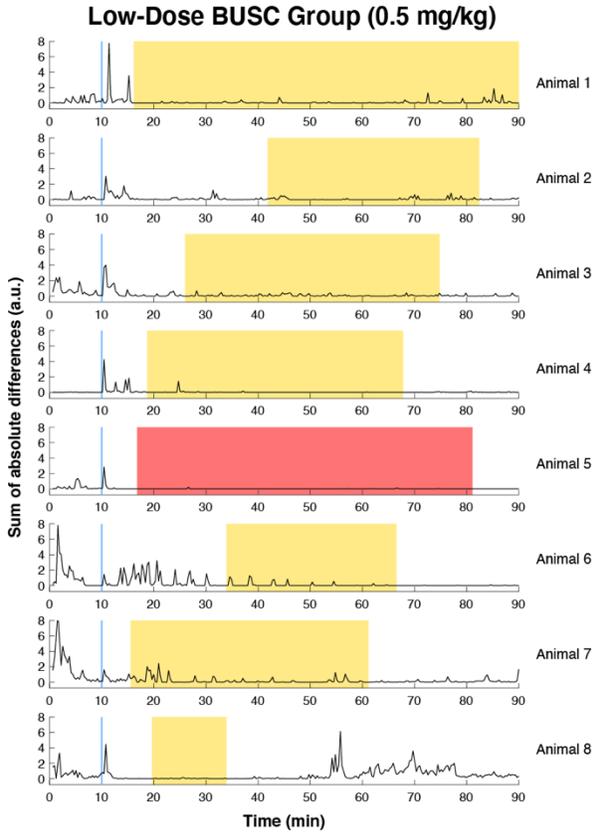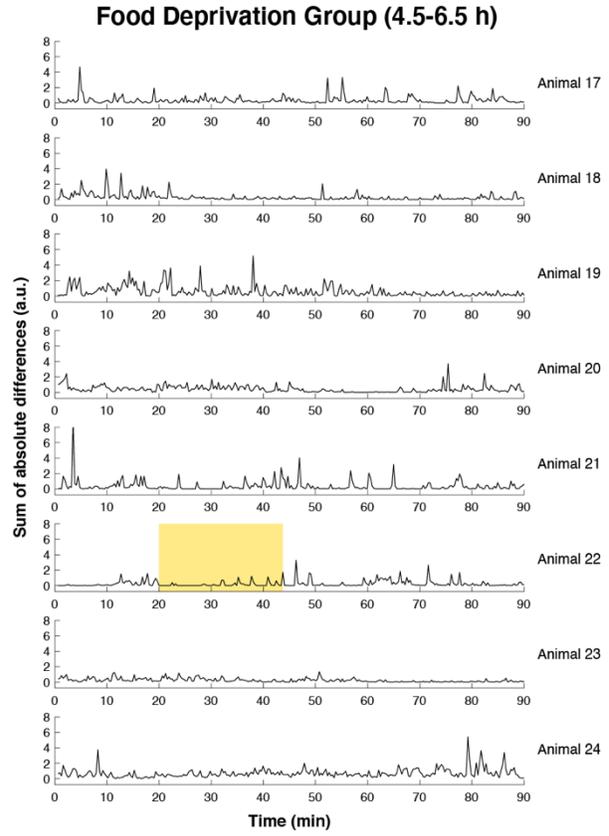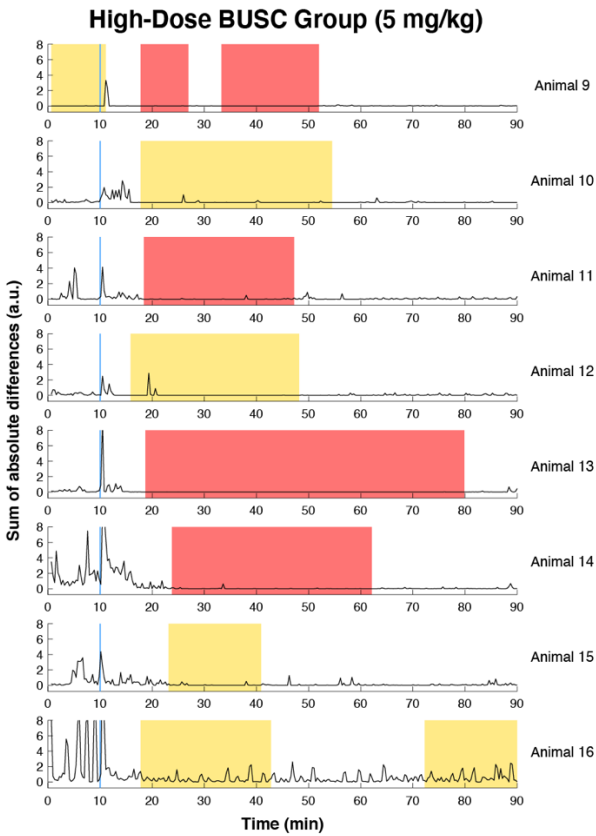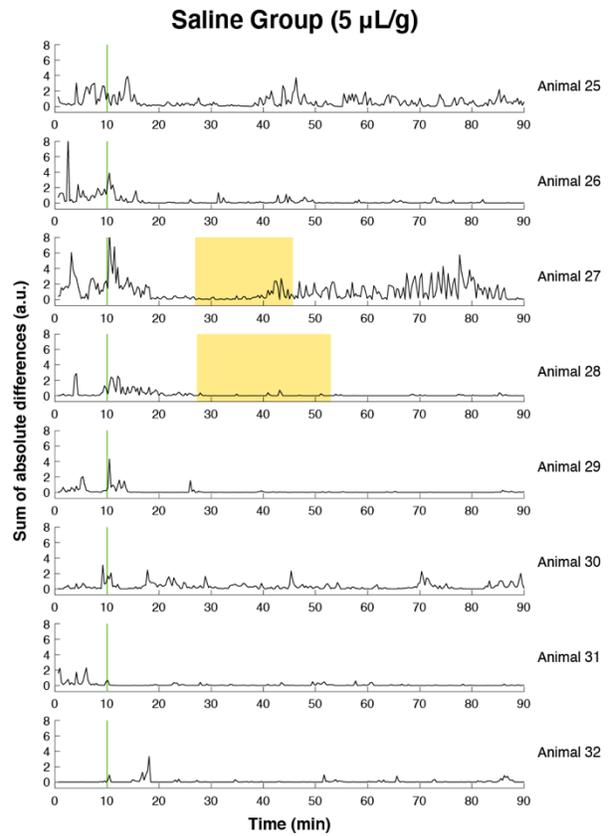

Figure S4 – **Individual motion time-courses**. Estimated motion time-courses obtained from both slices of each animal, depicting the sum of absolute differences between consecutive images for 19 sec intervals throughout the acquisition from significantly moving pixels, i.e. pixels whose absolute difference was higher than 20% of the highest difference, at each time instant. Yellow areas represent time intervals with reduced but residual peristalsis (classification = 2) and red areas represent time intervals where peristalsis was strongly reduced (classification = 1), both defined in consensus by the two readers. Blue lines indicate i.p. injection of a BUSC bolus whereas green lines represent i.p. injection of saline.

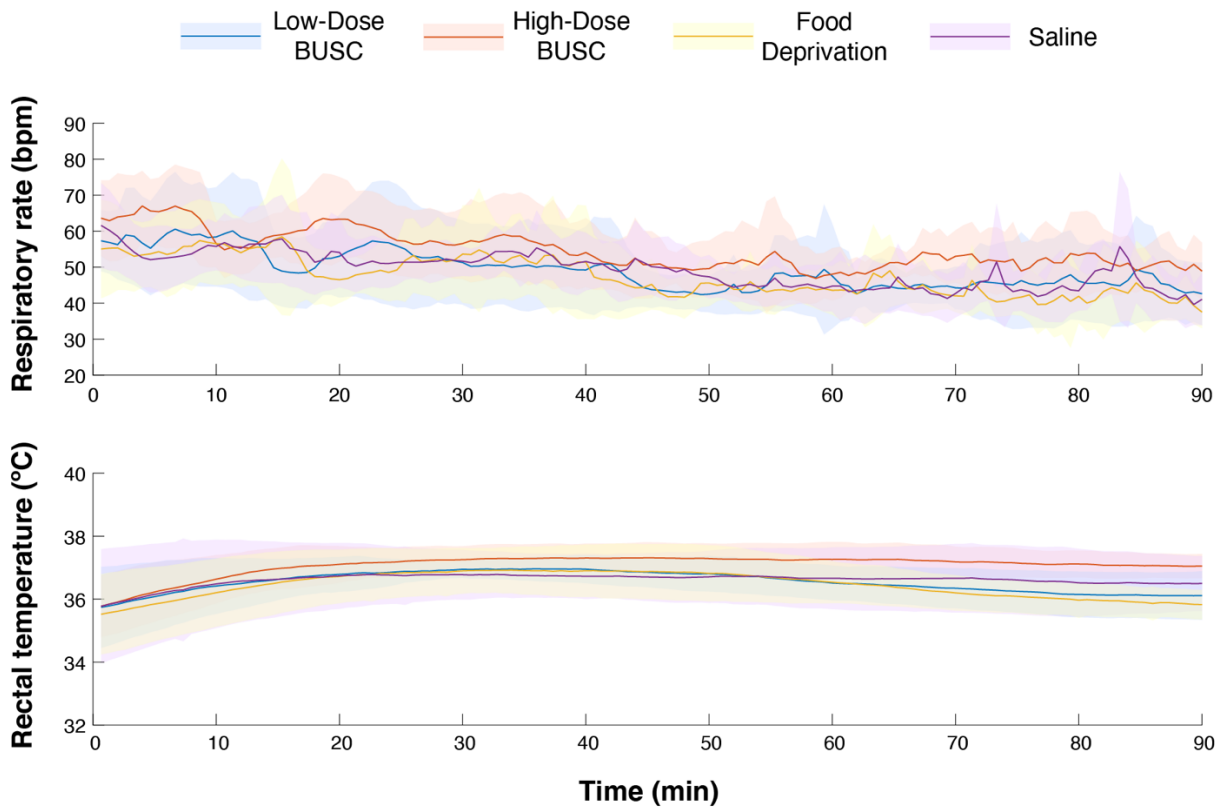

Figure S5 – **Group-averaged vital signs.** Respiratory rate and rectal temperature values recorded throughout MRI acquisition and averaged between animals of each group. Shaded areas represent the standard deviations of these measures.

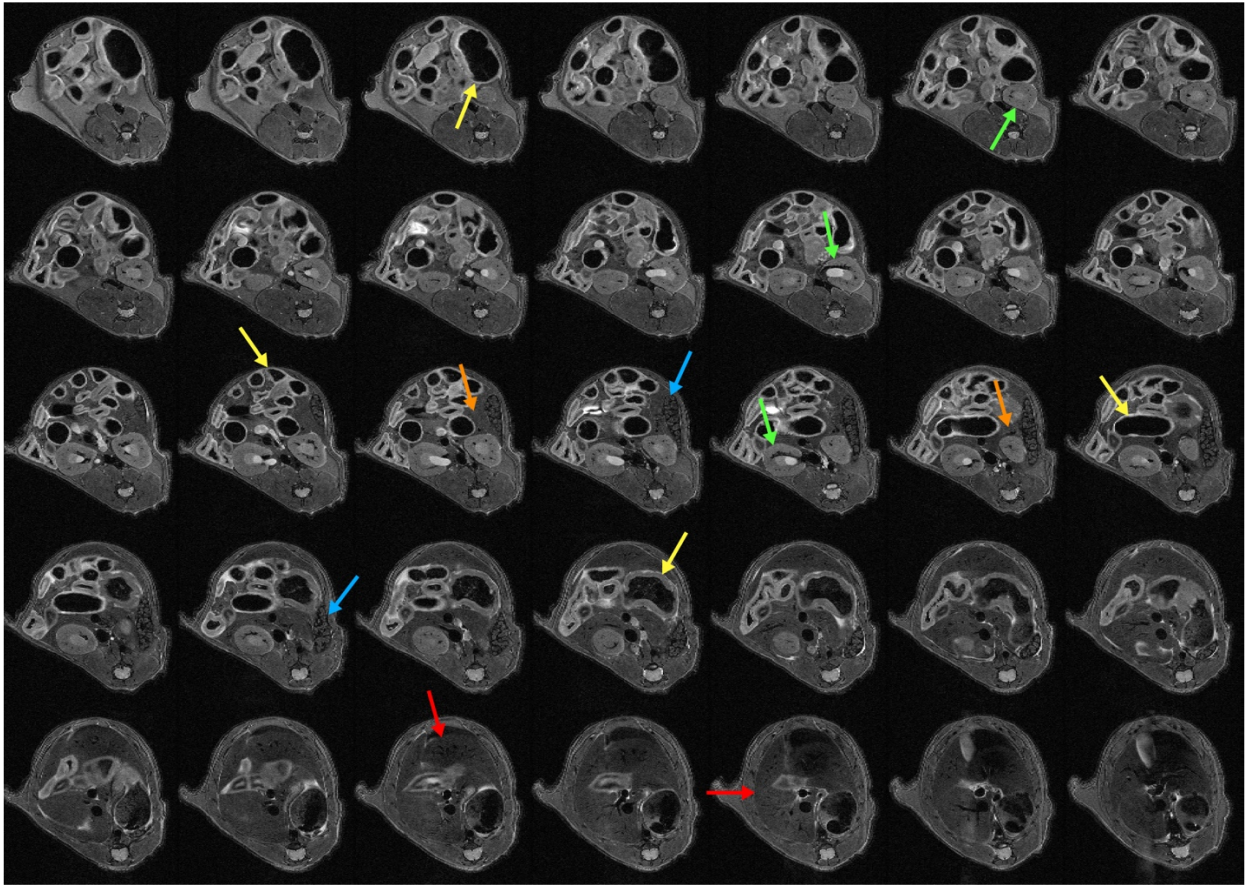

Figure S6 – **Axial images of the abdomen after BUSC injection**. High-resolution $T_2$-weighted images of a mouse abdomen acquired 6.5 min after BUSC i.p. injection, from caudal to cranial. Note the high definition and detail of abdominal organs: bowels (yellow arrows), kidneys (green arrows), pancreas (orange arrows), spleen (blue arrows) and liver (red arrows).

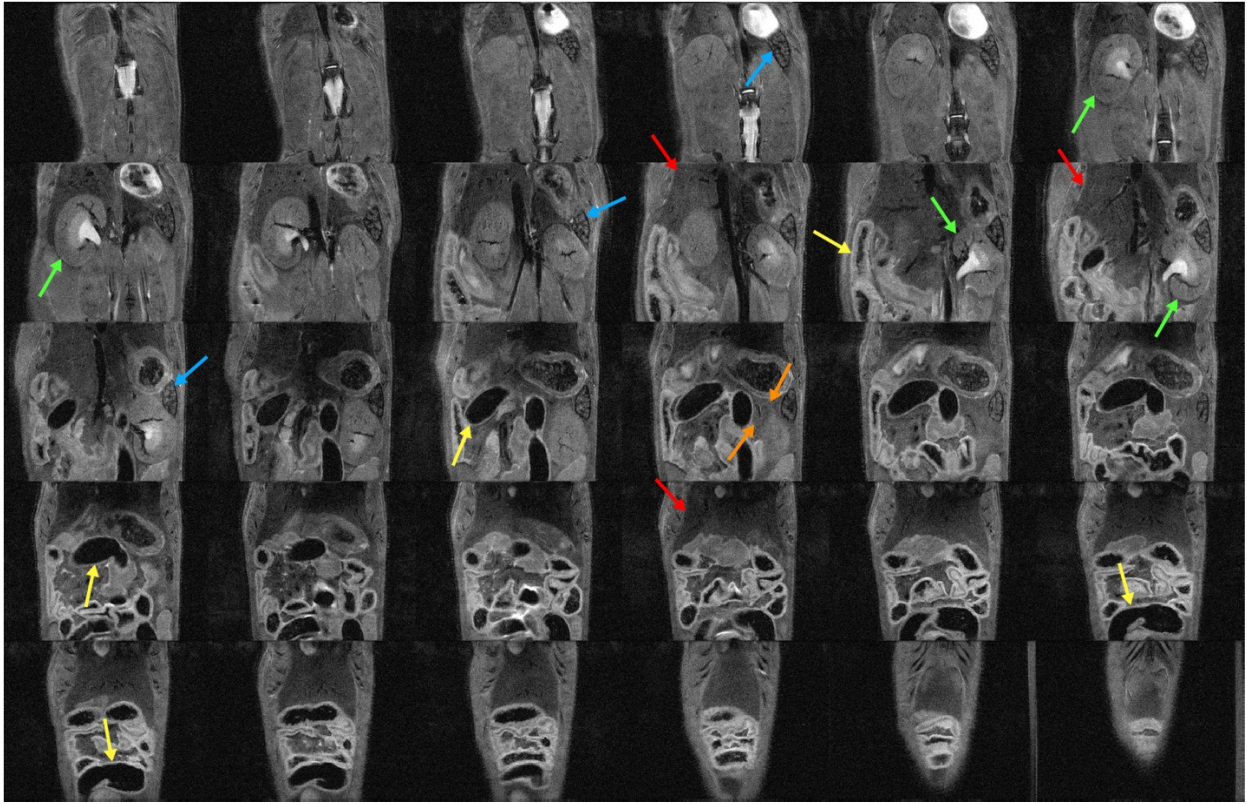

Figure S7 – **Coronal images of the abdomen after BUSC injection**. High-resolution $T_2$-weighted images of a mouse abdomen acquired 6.5 min after BUSC i.p. injection, from dorsal to ventral. Note the high definition and detail of abdominal organs: bowels (yellow arrows), kidneys (green arrows), pancreas (orange arrows), spleen (blue arrows) and liver (red arrows).

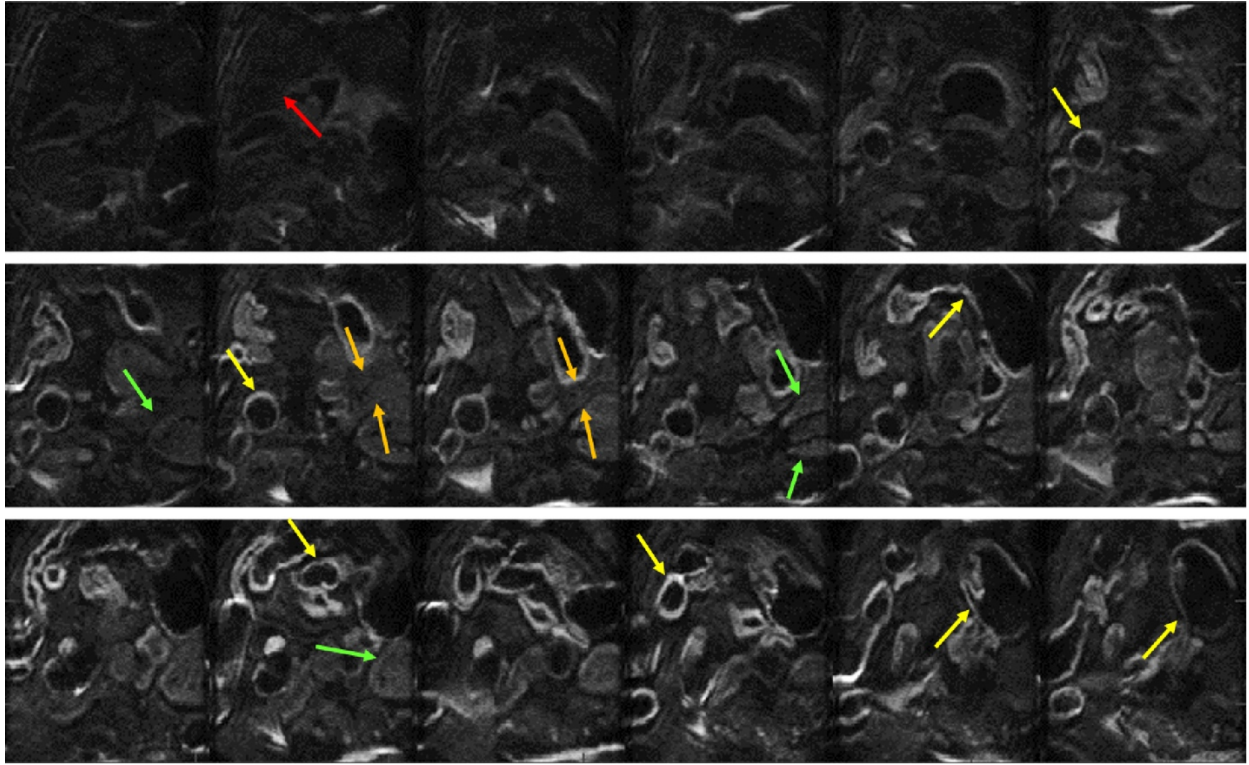

Figure S8 – **Axial images of the abdomen after BUSC injection.** DW (b1000) images of a mouse abdomen acquired 6.5 min after BUSC i.p. injection, from cranial to caudal. The abdominal organs are easily identified and well defined: bowels (yellow arrows), kidneys (green arrows), pancreas (orange arrows) and liver (red arrows).

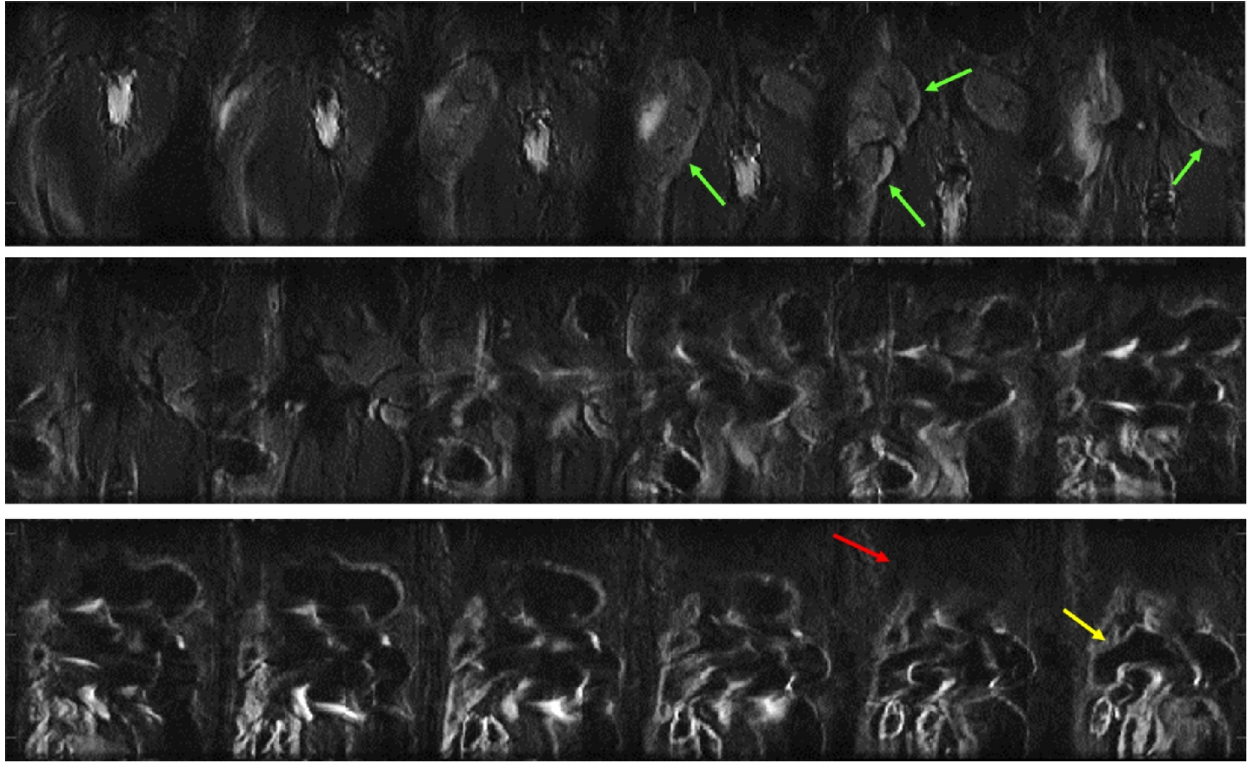

Figure S9 – **Coronal images of the abdomen after BUSC injection.** DW (b1000) images of a mouse abdomen acquired 6.5 min after BUSC i.p. injection, from dorsal to ventral. The abdominal organs are easily identified and well defined: bowels (yellow arrows), kidneys (green arrows) and liver (red arrows).

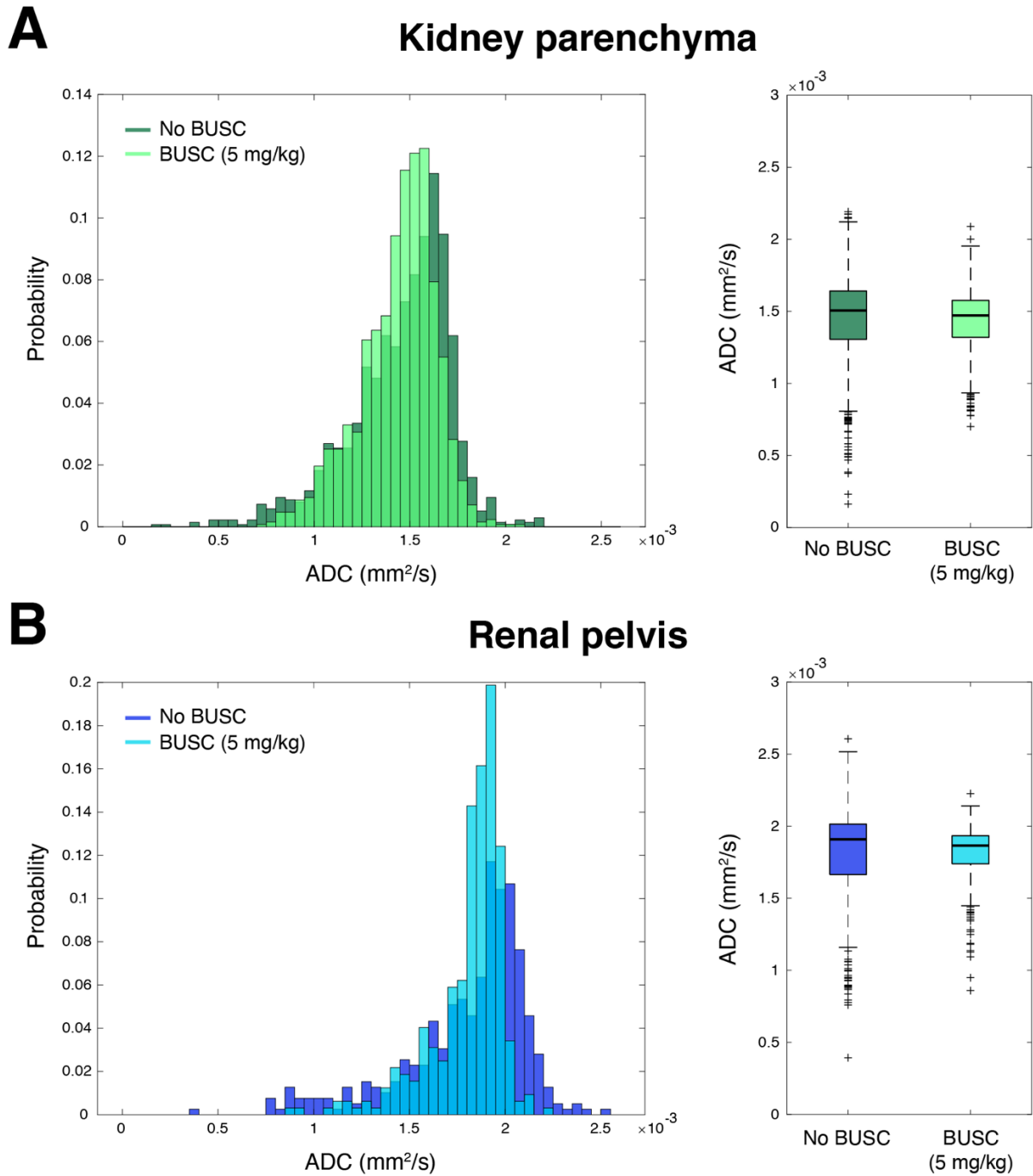

Figure S10 – **Histograms and boxplots of ADC values measured in kidney ROIs.** A higher variability of data can be observed in measurements performed before injecting BUSC, as represented by higher frequency of extreme values in both parenchymal (A) and renal pelvis (B) histograms and boxplots, when compared with measurements performed after injecting BUSC.

These measurements exemplify the applicability of BUSC for performing segmentation/quantification studies in the mouse abdomen, by reducing the rate of motion-derived artifactual measurements.

**SUPPLEMENTARY INFORMATION (METHODS)**

**Animal preparation**

Animals were reared in a temperature-controlled room and held under a 12h/12h light/dark regimen with *ad libitum* access to food and water.

Anesthesia was induced with 5% isoflurane (Vetflurane™, Virbac, France) mixed with oxygen-enriched (28%) air. Mice were then weighed, moved to the animal bed (Bruker BioSpin™, Germany) and isoflurane was reduced to 1.5-2.5%. Breathing rate and rectal temperature were monitored and recorded throughout the MRI sessions using a pillow sensor and an optic fiber probe (SA Instruments Inc., Stony Brook, USA), respectively. The readings were kept stable throughout the experiment via small adjustments to isoflurane levels (maintaining respiratory rates between 40-70 bpm) and a warm-water recirculating pad for body temperature control (maintaining temperatures between 35.0-37.0 ºC). Ophthalmic gel (Vidisic® Gel, Bausch+Lomb, Canada) was applied in the beginning of the experiment to prevent eye dryness.

**Characterization of BMR**

MRI protocol

For abdominal dynamic MRI, a FLASH sequence was acquired in n=32 animals with two 1-mm thick coronal slices positioned in the abdomen: TR/TE = 19/2 ms, flip angle = 20°, FOV = 25 x 25 $mm^2$, in-plane resolution = 120 x 120 $\mu m^2$, respiratory triggering, 284 repetitions, in a total duration of approximately 90 min (temporal resolution ~19 s).

Data analysis - Quantitative Analysis

Datasets were pre-processed in MATLAB® (MathWorks Inc., Natick, MA). Transient global signal increases caused by variations in the breathing rate were corrected by normalizing each single image to the sum of all of its voxel intensities (Figure S1A). Subsequently, images were corrected for sporadic stripe artifacts by calculating their 2D Fourier transform and eliminating spurious signal that appeared at high-frequency regions of the k-space. Affected regions of the k-space were first manually selected and then voxels in those regions were nulled whenever their values were higher than 1.5 times the standard deviation of their time-course. This value accurately selected all the strongly deviating k-space voxels that contributed to the artifacts, while still being permissible to voxels with weaker visual influence on the transformed images. In a single outlying case (Animal #16), a value of one standard deviation was chosen as the nulling criterion, because the stripe artifacts were less sporadic and more persistent during the entire scan. k-Space images were then 2D inverse-Fourier transformed to obtain the final corrected spatial images (Figure S1B).

**High-resolution MRI**

$T_2$-weighted images were acquired with Rapid Acquisition with Refocused Echoes (RARE), with the following parameters: TR/TE = 3500/25 ms, RARE factor = 8, in-plane resolution = 85 x 85 $\mu m^2$, FOV = 26 x 26 $mm^2$ (axial slices) or 20 x 26 $mm^2$ (coronal slices), number of slices = 42, slice thickness = 0.3 mm, slice gap = 0.25 mm, respiratory triggering; acquisition time: 49 min.

Diffusion-weighted images were acquired with a Stejskal-Tanner sequence with EPI: TR/TE = 2000/13.6 ms, 10 averages, 4 segments, double sampling, in-plane resolution = 150 x 150 $\mu m^2$, FOV = 16 x 18 $mm^2$ (if axial slices) or 18 x 18 $mm^2$ (if coronal slices), number of slices = 24 (axial slices) or 22 (coronal slices), slice thickness = 0.65 mm, slice gap = 0.2 mm, respiratory triggering, fat suppression, 14 directions of b = 1000 s/$mm^2$ (powder averaged), 2 acquisitions of b = 0; acquisition time: 30 min.